\documentclass[prl,aps,twocolumn]{revtex4}
\usepackage{amsmath,amssymb,graphicx}

\usepackage{pxfonts}
\usepackage{graphicx}
\begin{document}

\title{Localization-Delocalization Transition of Indirect Excitons in Lateral Electrostatic Lattices}

\author{M. Remeika, J.C. Graves, A.T. Hammack, A.D. Meyertholen, M.M. Fogler, and L.V. Butov}
\affiliation{Department of Physics, University of California at San
Diego, La Jolla, CA 92093-0319}

\author{M. Hanson and A.C. Gossard}
\affiliation{Materials Department, University of California at Santa
Barbara, Santa Barbara, California 93106-5050}

\begin{abstract}
We study transport of indirect excitons in GaAs/AlGaAs coupled
quantum wells in linear lattices created by laterally modulated gate
voltage. The localization-delocalization transition (LDT) for
transport across the lattice was observed with reducing lattice
amplitude or increasing exciton density. The exciton interaction
energy at the transition is close to the lattice amplitude. These
results are consistent with the model, which attributes the LDT to
the interaction-induced percolation of the exciton gas through the
external potential. We also discuss applications of the lattice
potentials for estimating the strength of disorder and exciton
interaction.

\end{abstract}


\date{\today}

\maketitle

Transport of particles in periodic potentials is a basic problem, which concerns
a variety of systems extending from condensed matter systems with electrons in
ionic lattices to engineered systems such as photons in photonic crystals and
cold atoms in optical lattices. The particle localization and LDT are perhaps
the most interesting transport phenomena. Particular cases of the latter --- the
metal-insulator and superfluid-insulator transitions --- have been extensively
studied for electrons, photons, and cold atoms in lattices \cite{Lee85,
Wiersma97, Greiner02, Fertig05, Chin06, Roati08}.

Tunability of system parameters has been essential in studies of cold atoms in
optical lattices, allowing to probe localization of atoms with increasing
lattice amplitude \cite{Greiner02,Fertig05,Chin06,Roati08}. Here, we study a
condensed matter system with tunable parameters: excitons in electrostatic
lattices created by a gate voltage. In this system parameters of both the
lattice, e.g., the lattice amplitude, and the particles, e.g., the exciton
density, can be controlled.

\begin{figure}
\begin{center}
\includegraphics[width=8.5cm]{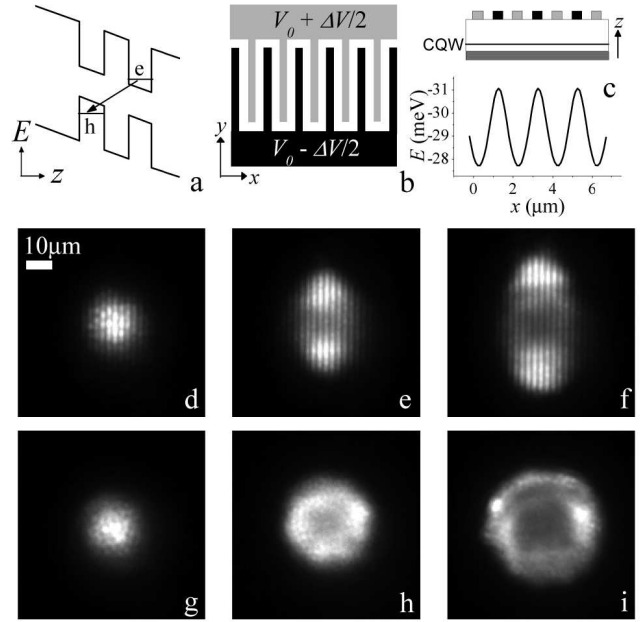}
\caption{(a) Energy band diagram of the CQW. (b,c) Schematic
electrode pattern. The applied base voltage $V_0$ realizes the
indirect regime while the voltage modulation $\Delta V$ controls the
lattice amplitude. Calculated lattice potential for indirect
excitons for $\Delta V = 1\,\text{V}$ is shown in (c). PL images of
indirect excitons for lattice amplitude (g-i) $\Delta V=0$
and (d-f) $\Delta V = 1.2\,\text{V}$ for excitation powers $P=0.2$ (d,g),
3.7 (e,h), and 12 (f,i) $\mu\,\text{W}$. $T=1.6\, \text{K}$, $\lambda_{ex}=633\,\text{nm}$,
and $V_0=3\,\text{V}$ for the data.}
\end{center}
\end{figure}

An indirect exciton in coupled quantum wells (CQW) is a bound state
of an electron and a hole in separate wells (Fig. 1a). Lifetimes of
indirect excitons exceed that of regular excitons by orders of magnitude
and they can travel over large distances before
recombination~\cite{Hagn95, Larionov00, Butov02, Voros05, Ivanov06,
Gartner06}. Also, due to their long lifetime, these bosonic
particles can cool to temperatures well below the quantum degeneracy
temperature $T_{\text{dB}}=2\pi \hbar^2 n / (m g k_{\text{B}})$
\cite{Butov01}. (In the studied CQW, excitons have the mass $m =
0.22 m_0$, spin degeneracy $g = 4$, and $T_{\text{dB}} \approx
3\,\text{K}$ for the density per spin $n / g =
10^{10}\,\text{cm}^{-2}$). Furthermore, indirect excitons in CQW
have a dipole moment $e d$, where $d$ is close to the distance
between the QW centers. This allows imposing external potentials
$E(x,y) = e d F_z(x,y) \propto V(x,y)$ for excitons using a
laterally modulated gate voltage $V(x,y)$, which creates a
transverse electric field $F_z(x,y)$ \cite{Hagn95, Zimmermann97,
Huber98, Krauss04, Hammack06, Chen06, Gartner06,
High07, High08}.

The lattice potential for indirect excitons $E(x)$ was created by interdigitated
gates. Base voltage $V_0=3\,\text{V}$ realized the indirect regime where
indirect excitons are lower in energy than direct excitons, while voltage
modulation $\Delta V$ controlled the lattice amplitude (Fig. 1b,c). Note that
in-plane electric field $F_r$ present near electrode edges can lead to exciton
dissociation \cite{Zimmermann97}. Therefore the CQW layers in our structure were
positioned closer to the homogeneous bottom electrode. This design suppresses
$F_r$ making the field-induced dissociation negligible \cite{Hammack06}. An
example of the calculated~\cite{Hammack06} (unscreened) $E(x)$ is shown in
Fig.~1(c). Zero energy corresponds to zero voltage, the 4 meV energy shift due
to binding energy of the indirect exciton is not shown. The potential modulation
is nearly sinusoidal $E(x) = E_{{0}} \cos^2 \left({q x}/\,{2} \right)$. Its
amplitude is $E_{{0}} = 3\,\text{meV}$ for $\Delta V = 1\,\text{V}$ and scales
linearly with $\Delta V$. The lattice period $2 \pi / q = 2\, \mu\text{m}$ is
determined by the electrode dimensions.

CQW structure was grown by MBE. $n^+$-GaAs layer with $n_{Si} =
10^{18}\,\text{cm}^{-3}$ serves as a homogeneous bottom electrode.
Semitransparent top electrodes were fabricated by magnetron sputtering a
$90\,\text{nm}$ indium tin oxide layer. CQW with $8\,\text{nm}$ GaAs QWs
separated by a $4\,\text{nm}$ Al$_{0.33}$Ga$_{0.67}$As barrier were positioned
100 nm above the $n^+$-GaAs layer within an undoped $1\, \mu\text{m}$ thick
Al$_{0.33}$Ga$_{0.67}$As layer. Excitons were photogenerated by a 633 nm HeNe or
786 nm Ti:Sapphire laser focused to a spot $\sim 10\, \mu\text{m}$ in diameter
in the center of the $150 \times 150\, \mu\text{m}$ lattice. Exciton density was
controlled by the excitation power. Photoluminescence (PL) images of the exciton
cloud were captured by a CCD with a filter $800 \pm 5\, \text{nm}$ covering the
spectral range of the indirect excitons. The diffraction limited spatial
resolution was $1.5\, \mu\text{m}$ (N.A.=0.28). The spectra were measured using a
spectrometer with resolution $0.18\, \text{meV}$.

Figure 1 shows images of the exciton cloud at zero ($\Delta V =
E_{{0}} = 0$) and finite ($\Delta V = 1.2\,\text{V}$, $E_{{0}} =
3.7\,\text{meV}$) lattice amplitude for different excitation powers
$P$. At low $P$ (Fig.~1d,g), the cloud profile essentially coincides
with the laser excitation spot. This indicates that excitons do not
travel beyond the excitation spot, i.e., they are localized. On the
contrary, at high $P$ (Fig. 1e,f,h,i), the excitons spread beyond
the excitation spot indicating that they are delocalized. The LDT
occurs both with and without the lattice. While the cloud is
practically symmetric at $\Delta V = 0$ (Fig. 1h,i), at finite
$\Delta V$ it is compressed in the lattice direction (Fig. 1e,f).

\begin{figure}
\begin{center}
\includegraphics[width=8.5cm]{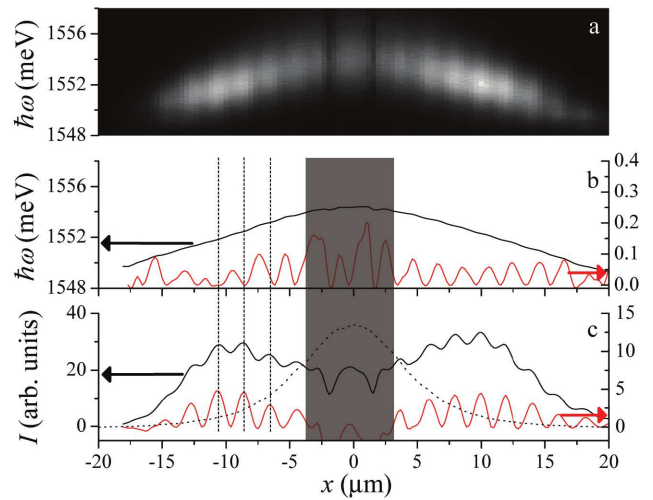}
\caption{(Color online) (a) The emission image in energy--$x$
coordinates for the lattice with $\Delta V = 1.2\,\text{V}$
($E_{{0}}=3.7\,\text{meV}$). The image was measured at the center $y =
0$ of the exciton cloud and integrated over $\Delta y =
1.5\,\mu\text{m}$. The corresponding (b) energy and (c) intensity
profiles (black, left scale). The same profiles with subtracted smooth
background are used to present the modulations in energy and intensity
(red, right magnified scale). The energy minima correspond to
intensity maxima. The PL linewidth does not exceed 2 meV that is
characteristic of excitonic emission. The dotted line shows the
profile of the laser excitation spot.  The shaded area contains two
deep intensity minima caused by a defect in the spectrometer slit. $T
= 1.6\,\text{K}$, $P = 35\,\mu\text{W}$,
$\lambda_{ex}=633\,\text{nm}$, and $V_0 = 3\,\text{V}$ for the data.}
\end{center}
\end{figure}

The lattice potential also causes periodic modulations of PL
characteristics. Figure 2 presents a PL image in energy--$x$
coordinates in the delocalized regime. Both the integrated PL
intensity $I(x)$ and the average PL energy $\hbar \omega(x)$ show
small modulation at the lattice period superimposed on a smoothly
varying profile (Fig. 2b,c). To demonstrate the modulations more
clearly, we subtracted the smooth component and plot the remainder
on a magnified scale in Fig.~2b,c. Minima in energy correspond
to the maxima in intensity. We define the amplitude of energy
modulation as the difference between adjacent maxima and minima
$\delta\omega = \omega_{\text{max}} - \omega_{\text{min}}$. Figure 2
shows that $\hbar \delta\omega$ is much smaller than the lattice
amplitude $E_{{0}} = 3.7\, \text{meV}$ that is discussed below. No
intensity or energy modulation was observed at $\Delta V = 0$. This
indicates that the modulations in question are not due to the
partial light absorption in the top electrodes.

The PL intensity has a maximum along a ring in the regime of
delocalized excitons (Figs. 1e,f,h,i, 2c). This so-called inner ring
was previously observed in PL patterns of indirect excitons without
lattices~\cite{Butov02}. It was explained in terms of exciton
transport and cooling~\cite{Butov02, Ivanov06}. The inner ring
effect persists in the lattice (Figs. 1e,f, 2c).

\begin{figure}
\begin{center}
\includegraphics[width=8.5cm]{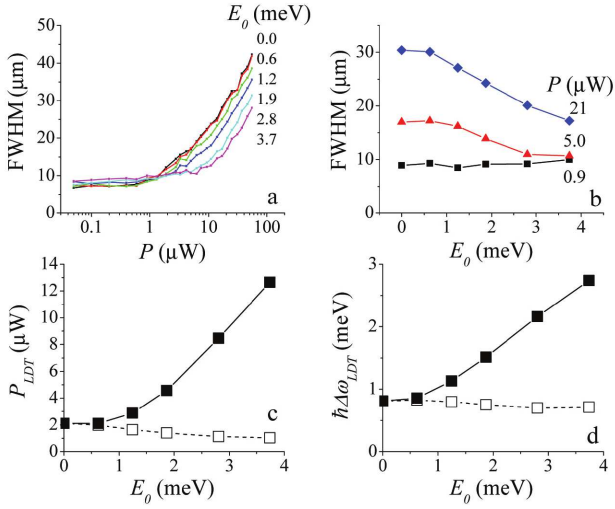}
\caption{The FWHM of the exciton cloud across the lattice (a) vs. the
excitation power $P$ for lattice amplitudes $E_{{0}} = 0$, $0.6$,
$1.2$, $1.9$, $2.8$, $3.7\, \text{meV}$ ($\Delta V = 0$, $0.2$,
$0.4$, $0.6$, $0.9$, $1.2\, \text{V}$) and (b) vs.
$E_{{0}}$ for $P = 0.9$, $5$, $21\, \mu\text{W}$. (c) The excitation
power at the transition from the localized to delocalized regime
$P_{LDT}$ as a function of $E_{{0}}$. (d) The interaction energy in
the center of the exciton cloud at the transition from the localized
to delocalized regime $\hbar \Delta\omega_{\text{LDT}}$ as a
function of $E_{{0}}$. Filled (open) squares in (c,d) present the
data for the exciton transport across (along) the lattice. $T =
1.6\,\text{K}$, $\lambda_{ex}=786\, \text{nm}$, and $V_0=3\,
\text{V}$ for the data.}
\end{center}
\end{figure}

The full width at half maximum of the exciton cloud in the
$x$-direction is plotted in Fig. 3a,b. Initially, it is practically
independent of $P$ but then starts to grow as $P$ increases. We
define the excitation power at the transition $P_{\text{LDT}}$ as
the point where the extrapolation of this growth to small $P$
becomes equal to the low-$P$ constant. At the LDT, the exciton cloud
starts to spread beyond the excitation spot and the cloud extension
changes from constant to increasing with $P$. Figure 3a shows that
the transition is smooth. This yields $P_{\text{LDT}} \approx
2\,\mu\text{W}$ for low lattice amplitudes $E_{{0}} \lesssim
1\,\text{meV}$. At higher $E_{{0}}$ the LDT for the $x$-direction
shifts to higher excitation powers with increasing lattice amplitude
(Fig. 3a,c). The LDT was also observed with reducing lattice
amplitude, see the data for $P=5\, \mu$W (Fig. 3b). Note that
excitons remain localized at lower $P=0.9\, \mu$W and delocalized at
higher $P=21\, \mu$W for all $E_{{0}}$ in Fig. 3b.

The exciton transport along the $y$-direction is only weakly
effected by the lattice (Fig. 1d-i). The LDT for this direction
shifts slightly to lower excitation powers with increasing $E_{{0}}$
(Fig. 3c).

The smooth component of the $\hbar\omega(x)$ also exhibits an
interesting behavior. It increases with increasing exciton density,
both with increasing $P$ or reducing $\vert x \vert$. Let
$\Delta\omega_{\text{LDT}}$ denote the difference between the value of $\omega$ at the
lowest $P$ and at the LDT at $x=0$. Figure~3d presents the
dependence of $\hbar \Delta\omega_{\text{LDT}}$ on $E_{{0}}$. We see
that it is finite at $E_{{0}} = 0$. At large $E_{{0}}$ we observe a
remarkable relation 
\begin{equation}\label{eqn:omega_LDT}
\hbar \Delta\omega_{\text{LDT}} \approx E_{0}\,,
\end{equation}
which is crucial to our interpretation of the mechanism of the LDT,
see below.

Let us now discuss a simple model that attributes the observed LDT
to the interaction-induced percolation of exciton gas through the
total external potential $E_{\text{tot}}(\mathbf{r})$, which is the
sum of the periodic lattice potential $E(x)$ and the random
potential $E_{\text{rand}}(\mathbf{r})$ due to disorder. The latter
is an intrinsic feature of solid state materials. It forms mainly
due to QW width and alloy fluctuations in the structure.

The idea is illustrated in Fig.~4 for the case of no disorder,
$E_{\text{rand}}(\mathbf{r}) \equiv 0$. If the local exciton density
$n(x)$ is low, it is concentrated in the minima of the potential
$E(x)$. The crests are nearly depleted. As a result, the exciton
transport from one period of the lattice to the next through
thermal activation or quantum tunneling is exponentially slow. As the average
density increases and reaches a certain threshold --- ``percolation
point'' --- the crests become populated, which permits a faster
exciton transport, i.e., the observed delocalization. This scenario
naturally leads to Eq.~\eqref{eqn:omega_LDT}, see Fig.~4b, where the
middle curve corresponds to the percolation point. It also explains
why $P_{\text{LDT}}$ increases as $E_{{0}}$ goes up, see Fig.~3d.

Adding disorder does not modify this picture greatly as long as
$E_{{0}}$ remains larger than the characteristic amplitude of
$E_{\text{rand}}$. Otherwise, the percolation is determined by the
random potential~\cite{Lugan07, Chen08}, so that the dependence of $P_{\text{LDT}}$ and
$\Delta\omega_{\text{LDT}}$ on $E_{{0}}$ saturates. The
saturation point gives the estimate of $E_{\text{rand}}$. From
Fig.~3d, we find $E_{\text{rand}} \sim 0.8\,\text{meV}$.
This number is comparable to the PL linewidth at low densities,
suggesting that the disorder is responsible
for both of these energy scales.

\begin{figure}[b]
\begin{center}
\includegraphics[width=8.5cm]{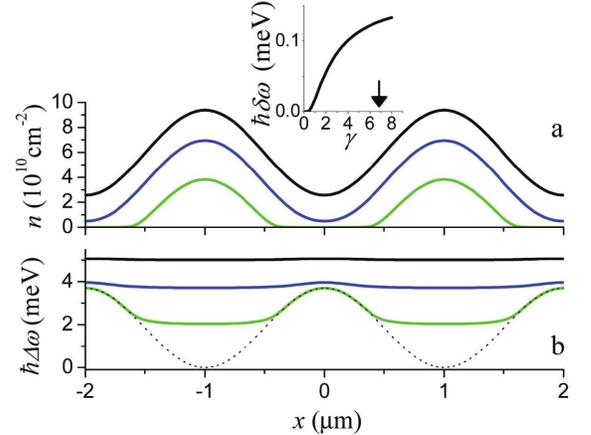}
\end{center}
\caption{(a) Exciton density for (top to bottom) $\zeta = 5$, $3.7$,
and $2.5\,\text{meV}$. The first of these corresponds to data at $x
\sim 10\,\mu\text{m}$ in Fig.~2. $k_B T = 0.15\,\text{meV}$,
$E_{{0}} = 3.7\,\text{meV}$, and $\gamma = 2.3$ for all curves. (b) Lattice potential
$E(x)$ and the PL
energy shift $\hbar\Delta\omega(x)$ for the same set of parameters.
Inset: Modulation $\delta \omega = \omega_{\text{max}} -
\omega_{\text{min}}$ of the PL energy as a function of the
interaction strength $\gamma$. The experimental $\delta\omega$ corresponds to $\gamma
\approx 2.3$. The value of $\gamma$ predicted by the ``capacitor''
formula~\cite{Yoshioka90, Zhu95, Lozovik96, Ivanov02} is indicated
by the arrow.}
\label{fig:n_lattice}
\end{figure}

To further develop this model, we make the following simplifying
assumptions: 1) $E_{\text{rand}} \equiv 0$, while $E(x)$ can be considered
slowly varying, 2) excitons reach a quasi-equilibrium state with
local chemical potential $\mu(x)$ and temperature $T(x)$, which are
also slowly varying, 3) exciton interaction is local (dipolar tails,
see below, are neglected).
Under these assumption, we obtain
\begin{align}
\zeta(x) &\equiv \mu(x) + E(x) \simeq \text{const}\,,
\label{eqn:zeta}\\
n(x) &\simeq \int\limits_0^\infty
\frac{g \nu_1 d \epsilon}
     {\exp \big[\big(\epsilon + \text{Re}\,\Sigma(\epsilon,x) - \mu(x)\big)\,/\, k_B T(x)\big] - 1}\,,
\label{eqn:n_I}
\end{align}
where $\zeta$ is the electrochemical potential, $\nu_1 = m / (2 \pi \hbar^2)$
is the density of states per spin species, and
$\Sigma(\epsilon)$ is the self-energy
(in the uniform state of the same $n$).

To find the equation for the PL energy shift $\hbar\Delta\omega$ we
take advantage of the smallness of $Q$, the range of in-plane
momenta collected by our optical system. It is given by $Q \equiv 2
\pi\, \text{N.A.} / \lambda \approx (0.45\,\mu\text{m})^{-1}$, which
is indeed small. In this case $\hbar \Delta\omega(x) = \text{Re}\,
\Sigma(0,x) + E(x)$.

To complete the system of equations we need a formula for
$\Sigma(\epsilon,x)$. This self-energy
is due to the exciton interaction. At
large $r$ the interaction is known to be dominated by dipole
repulsion $e^{2}d^{2} / (\kappa r^{3})$.
When excitons approach each other, the interaction potential
becomes complicated. What appears to be certain is that for $d = 12\,\text{nm}$
in our experiment the exciton interaction remains strictly
repulsive~\cite{Schindler08, Meyertholen08}, and so $\text{Re}\,\Sigma(0)$
increases with density: $\text{Re}\,\Sigma(0) = t n$. The growth of
$\text{Re}\,\Sigma(0)$ with $n$ implies an increase of the PL energy
$\hbar\omega$, which is observed experimentally. The calculation of function $t
= t(n, T) > 0$ remains a challenging open problem~\cite{Yoshioka90, Zhu95,
Lozovik96, Laikhtman01, Ivanov02, Schindler08}. Therefore, we treat $t$ as
a phenomenological constant. We also assume that
$\text{Re}\,\Sigma(\epsilon) \simeq \text{Re}\,\Sigma(0)$, which
is reasonable for short-range interactions.
Substituting $\Sigma = t n$ into Eq.~\eqref{eqn:n_I}, after some algebra
we get
\begin{equation}\label{eqn:n_II}
 \exp\left(-\frac{n}{k_B T N \nu_1}\right) +
 \exp\left(\frac{\mu \nu_1 - g n}{k_B T \nu_1}\right) = 1\,,
 \:\:\:
 \mu = \zeta - E(x)\,,
\end{equation}
where the dimensionless parameter $\gamma = t \nu_1$ characterizes the strength of
the interaction. Given $\zeta$, $T$, and $\gamma$, Eq.~\eqref{eqn:n_II} can be solved
numerically for each $E(x)$. The results for $n(x)$ and $\Delta\omega(x)$ are
shown in Fig.~4a,b.

The exciton interaction results in screening of the lattice
potential at points where local density is not small. Because of
this screening, the amplitude of the PL energy modulation
$\hbar\delta\omega$ is much smaller than $E_{{0}}$. Consider points
$|x| \approx 10\,\mu\text{m}$, which are midway between the center
and the edge of the exciton cloud in Fig.~2, on the aforementioned
inner ring. (Here the exciton temperature is close to the sample
temperature $T = 1.6\,\text{K}$ \cite{Ivanov06}.) From Fig.~2b we
see that $\hbar\delta\omega(x) \approx 0.07\,\text{meV}$, more than
an order of magnitude smaller than $E_{{0}}=3.7\, \text{meV}$.

Using the above equations we calculated $\delta\omega$ as a function of the
adjustable parameter $\gamma$, cf.~the inset of Fig.~4. The experimental value of
$\hbar\delta\omega$ gives a rough estimate $\gamma \approx
2.3$. In comparison, the mean-field Hartree approximation~\cite{Yoshioka90,
Zhu95, Lozovik96, Ivanov02} yields the so-called ``plate capacitor'' formula
$\gamma_\text{cap} = (2 d/a_e)(m/m_e) \approx 7$, where $a_e = 10\,\text{nm}$ is the
electron Bohr radius, see Fig.~4. The reduction of the interaction constant
compared to the $\gamma_\text{cap}$ can be due to correlation
effects~\cite{Laikhtman01, Schindler08}. A systematic analysis of $\gamma$ remains a
problem for future research.

This work is supported by the DOE Grant ER46449. A.D.M. and M.M.F.
are supported by the NSF Grant DMR-0706654. We thank A.~L.~Ivanov,
L.~S.~Levitov, L.~J.~Sham, and C.~J.~Wu for discussions.

\end{document}